
\magnification \magstep1
\raggedbottom
\openup 4\jot
\voffset6truemm
\headline={\ifnum\pageno=1\hfill\else
\hfill {\it Mixed boundary conditions in Euclidean
quantum gravity} \hfill \fi}
\def\cstok#1{\leavevmode\thinspace\hbox{\vrule\vtop{\vbox{\hrule\kern1pt
\hbox{\vphantom{\tt/}\thinspace{\tt#1}\thinspace}}
\kern1pt\hrule}\vrule}\thinspace}
\rightline {June 1995, DSF preprint 95/23}
\centerline {\bf MIXED BOUNDARY CONDITIONS IN}
\centerline {\bf EUCLIDEAN QUANTUM GRAVITY}
\vskip 1cm
\centerline {Giampiero Esposito$^{1,2}$
and Alexander Yu Kamenshchik$^{3}$}
\vskip 1cm
\centerline {\it ${ }^{1}$Istituto Nazionale di Fisica Nucleare,
Sezione di Napoli}
\centerline {\it Mostra d'Oltremare Padiglione 20,
80125 Napoli, Italy;}
\centerline {\it ${ }^{2}$Dipartimento di Scienze Fisiche}
\centerline {\it Mostra d'Oltremare Padiglione 19,
80125 Napoli, Italy;}
\centerline {\it ${ }^{3}$Nuclear Safety Institute, Russian
Academy of Sciences}
\centerline {\it Bolshaya Tulskaya 52, Moscow 113191, Russia.}
\vskip 1cm
\noindent
{\bf Abstract.} This paper studies a new set of mixed
boundary conditions in Euclidean quantum gravity. These
involve, in particular, Robin boundary conditions on the
perturbed 3-metric and hence lead, by gauge invariance,
to Robin conditions on the whole ghost 1-form. The corresponding
trace anomaly is evaluated in the case of flat Euclidean
4-space bounded by a 3-sphere. In general, this anomaly
differs from the ones
resulting from other local or non-local boundary conditions
studied in the recent literature.
\vskip 1cm
\noindent
PACS: 0370, 0460, 9880
\vskip 100cm
\leftline {\bf 1. Introduction}
\vskip 1cm
\noindent
Since the main problem in quantum cosmology is to find a
suitable set of boundary conditions which completely determine
the path integral for the
quantum state of the universe [1-2], the search for such
boundary conditions is of crucial importance in
Euclidean quantum gravity [3]. For this purpose, it has been
found in [4] that, following earlier work in [5], a set of
gauge-invariant boundary conditions for Euclidean quantum
gravity can be written as
$$
\Bigr[h_{ij}\Bigr]_{\partial M}=0
\eqno (1.1)
$$
$$
\left[{\partial h_{00}\over \partial \tau}+{6\over \tau}
h_{00}-{\partial \over \partial \tau}(g^{ij}h_{ij})
+{2\over \tau^{2}}h_{0i}^{\; \; \; \mid i}
\right]_{\partial M}=0
\eqno (1.2)
$$
$$
\left[{\partial h_{0i}\over \partial \tau}
+{3\over \tau}h_{0i}-{1\over 2}h_{00 \mid i}
\right]_{\partial M}=0
\eqno (1.3)
$$
$$
\Bigr[\varphi_{0}\Bigr]_{\partial M}=0
\eqno (1.4)
$$
$$
\Bigr[\varphi_{i}\Bigr]_{\partial M}=0 .
\eqno (1.5)
$$
With our notation [4], $g$ is the flat background 4-metric,
$h$ is its perturbation, $\tau$ is the Euclidean-time
coordinate, which becomes a radial coordinate if flat
4-space is bounded by a 3-sphere (see section 3).
Moreover, $\varphi_{\mu}$
is the ghost 1-form, with normal component $\varphi_{0}$ and
tangential components $\varphi_{i}$, and the stroke denotes
covariant differentiation tangentially with respect to the
3-dimensional Levi-Civita connection of the boundary.
We are interested in the Faddeev-Popov formalism for quantum
amplitudes. Thus, we deal with a gauge-averaging term in
the Euclidean action, i.e.
$$
I_{g.a.}={1\over 32\pi G \alpha} \int_{M}\Phi_{\nu}\Phi^{\nu}
\sqrt{{\rm det} \; g} \; d^{4}x
\eqno (1.6)
$$
where $\Phi_{\nu}$ is taken to be the de Donder functional
($\nabla$ being the 4-dimensional Levi-Civita connection
of the background) [4,6]
$$
\Phi_{\nu}^{dD}(h) \equiv \nabla^{\mu}
\Bigr(h_{\mu \nu}-{1\over 2}g_{\mu \nu}g^{\rho \sigma}
h_{\rho \sigma}\Bigr)
\eqno (1.7)
$$
and we consider a ghost $\eta_{\mu}$ and an anti-ghost
${\overline \eta}_{\mu}$. These ghosts reflect the invariance
of the classical theory under {\it gauge} transformations of metric
perturbations of the form
$$
{\widehat h}_{\mu \nu} \equiv h_{\mu \nu}
+{1\over 2}\nabla_{\mu}\varphi_{\nu}
+{1\over 2}\nabla_{\nu}\varphi_{\mu} .
\eqno (1.8)
$$
The boundary conditions for $\varphi_{\mu}$ are the same as
those for $\eta_{\mu}$ and ${\overline \eta}_{\mu}$ [7].
This property should not be surprising, since already in
the simpler case of Euclidean Maxwell theory the ghost
and anti-ghost obey the same boundary conditions imposed
on the scalar field occurring in the gauge transformations
of the potential (see Appendix A).
It is therefore sufficient to consider $\varphi_{\mu}$,
and then multiply by $-2$ the resulting contribution to the
amplitudes, since the ghosts are anti-commuting and
complex-valued.

The boundary conditions (1.1) reflect a natural choice in
the classical variational problem in general relativity,
where one often fixes at the boundary the intrinsic 3-geometry
[2,8]. Since, denoting by $K_{ij}$ the unperturbed
extrinsic-curvature tensor of the boundary
($K_{ij} \equiv {1\over 2}{\partial g_{ij}\over \partial \tau}
=\tau c_{ij}={1\over \tau}g_{ij}$,
where $c_{ij}$ is the metric on a unit
3-sphere), one has
$$
{\widehat h}_{ij} \equiv h_{ij}+{1\over 2}\varphi_{i \mid j}
+{1\over 2}\varphi_{j \mid i}+K_{ij}\varphi_{0}
\eqno (1.9)
$$
the invariance of (1.1) under (1.9) is guaranteed providing
(1.4)-(1.5) hold. Strictly, one has to require that
$\Bigr[\varphi_{i \mid j}\Bigr]_{\partial M}=0$, but this is
automatically satisfied by (1.5), at least in the case of a
3-sphere boundary. The remaining boundary conditions (1.2)-(1.3)
are obtained by setting to zero at the boundary the de Donder
functional defined in (1.7). Their invariance under (1.8) is
guaranteed providing (1.4)-(1.5) hold.

Note that the boundary conditions (1.1)-(1.5) are non-local
in that they cannot be written in terms of complementary
projection operators, and can instead be re-expressed as
integral equations on metric perturbations at the boundary.
The aim of our paper is to study in detail another relevant
choice of mixed boundary conditions in Euclidean
quantum gravity. The motivations of our analysis are as
follows.
\vskip 0.3cm
\noindent
(i) In classical general relativity one can also fix at the
boundary the trace of the extrinsic-curvature tensor (rather
than the induced 3-metric) [8]. This choice of boundary
conditions is also relevant for the theory of the quantum
state of the universe [1,9].
\vskip 0.3cm
\noindent
(ii) In one-loop quantum cosmology, one can also consider
Hawking's magnetic boundary conditions for quantum gravity,
which set to zero at the boundary the linearized magnetic
curvature, and hence the first derivatives with respect to
$\tau$ of the perturbed 3-metric (see section 7.3 of [2]).
\vskip 0.3cm
\noindent
(iii) The boundary conditions (1.1)-(1.5) are gauge-invariant
but non-local. By contrast, the Luckock-Moss-Poletti boundary
conditions studied in [4,6,7,10], i.e.
$$
\Bigr[h_{ij}\Bigr]_{\partial M}=0
\eqno (1.10)
$$
$$
\Bigr[h_{0i}\Bigr]_{\partial M}=0
\eqno (1.11)
$$
$$
\biggr[{\partial h_{00}\over \partial \tau}
+{6\over \tau}h_{00}-{\partial \over \partial \tau}
(g^{ij}h_{ij})\biggr]_{\partial M}=0
\eqno (1.12)
$$
$$
\Bigr[\varphi_{0}\Bigr]_{\partial M}=0
\eqno (1.13)
$$
$$
\biggr[{\partial \varphi_{i}\over \partial \tau}
-{2\over \tau}\varphi_{i}\biggr]_{\partial M}=0
\eqno (1.14)
$$
are local but not entirely
gauge-invariant, since (1.14) makes it impossible to preserve (1.10)
under the action of (1.9).
It therefore appears
that neither locality nor complete gauge invariance of the boundary
conditions are always respected in Euclidean quantum gravity.
The minimal requirement is
instead that the boundary conditions on the
ghost 1-form $\varphi_{\mu}$ should lead to gauge invariance of
at least a subset of the boundary conditions on metric
perturbations. The latter choice reflects a careful analysis
of well-posed variational problems in classical general
relativity, which usually fix at the boundary the 3-metric,
or its normal derivatives, or the trace of the extrinsic-curvature
tensor [8].

Section 2 derives in detail our new set of mixed boundary
conditions for the linearized gravitational field. Section 3
evaluates the corresponding trace anomaly by using
zeta-function regularization. Concluding remarks and open
problems are presented in section 4. Relevant details on
boundary conditions and $\zeta(0)$ calculations are
described in the appendices.
\vskip 1cm
\leftline {\bf 2. Mixed boundary conditions}
\vskip 1cm
\noindent
For a given choice of quantization technique
and background 4-geometry, the amplitudes of
quantum gravity may depend on
the boundary 3-geometry and on the boundary conditions, and
hence a large number of interesting problems can be studied
for various gauge-averaging terms in the Euclidean action.
For the reasons described in the introduction, we are interested
in the most general class of Robin boundary conditions on
perturbations of the 3-geometry. Thus, denoting by $\lambda$ a
real dimensionless parameter, we begin by requiring that
$$
\biggr[{\partial h_{ij}\over \partial \tau}
+{\lambda \over \tau}h_{ij}\biggr]_{\partial M}=0.
\eqno (2.1)
$$
The invariance of the boundary conditions (2.1) under the
transformations (1.9) is ensured by the following set of
Robin conditions on the ghost 1-form:
$$
\biggr[{\partial \varphi_{0}\over \partial \tau}
+{(\lambda+1)\over \tau}\varphi_{0}\biggr]_{\partial M}=0
\eqno (2.2)
$$
$$
\biggr[{\partial \varphi_{i}\over \partial \tau}
+{\lambda \over \tau}\varphi_{i}\biggr]_{\partial M}=0 .
\eqno (2.3)
$$
Note that (2.3) is a local equation which ensures the
validity of a more complicated condition, involving its
3-dimensional covariant derivative with respect to the
Levi-Civita connection of the boundary.

At this stage, the remaining set of boundary conditions on
metric perturbations can be obtained by bearing in mind that,
in the case of flat Euclidean 4-space with a 3-sphere
boundary, one has from (1.8)
$$
{\widehat h}_{00}=h_{00}+{\partial \varphi_{0}\over \partial \tau}
\eqno (2.4)
$$
$$
{\widehat h}_{0i}=h_{0i}+{\partial \varphi_{i}\over
\partial \tau}-{2\over \tau}\varphi_{i} .
\eqno (2.5)
$$
Thus, if we require
$$
\Bigr[h_{00}\Bigr]_{\partial M}=0
\eqno (2.6)
$$
this boundary condition is gauge-invariant providing one sets
$\lambda=-1$ in (2.2), and if we require
$$
\Bigr[h_{0i}\Bigr]_{\partial M}=0
\eqno (2.7)
$$
this boundary condition is gauge-invariant providing one
sets $\lambda=-2$ in (2.3). Of course, the whole set of
boundary conditions (2.1)-(2.3), (2.6)-(2.7)
does not involve complementary projection
operators and is not gauge-invariant. Nevertheless, they seem to
correspond to the most general form of quantum boundary-value
problem which admits as a particular case (for $\lambda=0$)
the boundary conditions of section 7.3 of [2] on
transverse-traceless perturbations, and such that the boundary
conditions on the perturbed 3-geometry are always
gauge-invariant. If $\lambda$ vanishes,
they also provide a generalization to pure
gravity of the electric boundary conditions for Euclidean
Maxwell theory, whilst the Barvinsky boundary conditions
(1.1)-(1.5) generalize the magnetic boundary conditions
for Maxwell theory (see appendix A).
\vskip 1cm
\leftline {\bf 3. $\zeta(0)$ value}
\vskip 1cm
\noindent
The boundary conditions (2.1)-(2.3) and (2.6)-(2.7) are
now applied to evaluate the trace anomaly in quantum cosmology,
in the case of flat Euclidean 4-space bounded by a 3-sphere
[2,4,10,11]. This may be regarded as the first step towards the
analysis of curved backgrounds with boundary, or as a case
relevant for quantum cosmology in the limit of small
3-geometries [11].

The trace anomaly is expressed through the value at the origin
of the generalized zeta-function [12] obtained from the
eigenvalues of the elliptic operators acting on metric
perturbations and ghosts. Since the necessary formalism is
developed and applied in detail in [4,6,13,14], we limit
ourselves to a brief outline of the lengthy calculations
involved in our analysis. First, one expands on a family of
3-spheres centred on the origin the metric perturbations
$h_{00},h_{0i},h_{ij}$ and the components
$\varphi_{0},\varphi_{i}$ of the ghost 1-form. One then finds
10 sets of perturbative modes, 7 for $h_{\mu \nu}$ and 3
for $\varphi_{\mu}$, which give rise to 8 contributions to
$\zeta(0)$ resulting from transverse-traceless modes,
scalar-type perturbations, vectorlike perturbations, a finite
number of scalar modes, a decoupled
vector mode, plus scalar ghost modes, vector ghost modes and a
decoupled ghost mode. Decoupled modes belong to finite-dimensional
subspaces, and multiply scalar harmonics, or longitudinal
harmonics, or transverse harmonics.

The contributions of all modes to $\zeta(0)$ are
obtained by combining their uniform asymptotic expansions
as both the eigenvalues and the order tend to $\infty$ (the
modes being linear combinations of Bessel functions) with their
limiting behaviour as the eigenvalues tend to 0 and the order
tends to $\infty$. In the light of the boundary conditions of
section 2, the 8 contributions to $\zeta(0)$ are found to be
(see appendix B)
$$
\zeta(0)_{\rm transverse-traceless \; modes}
={112\over 45}+3\lambda-\lambda^{2}-{1\over 3}\lambda^{3}
\eqno (3.1)
$$
$$
\zeta(0)_{\rm scalar \; modes}
= {824\over 45}-2\lambda-\lambda^{2}
-{1\over 3}\lambda^{3}
\eqno (3.2)
$$
$$
\zeta(0)_{\rm vector \; modes}
={434\over 45}- 2\lambda - \lambda^{2}
-{1\over 3}\lambda^{3}
\eqno (3.3)
$$
$$
\zeta(0)_{\rm decoupled \; scalar \; modes}=-19
\eqno (3.4)
$$
$$
\zeta(0)_{\rm decoupled \; vector \; mode}=-{21\over 2}
\eqno (3.5)
$$
$$
\zeta(0)_{\rm scalar \; ghost \; modes}=-{59\over 45}
+2\lambda+{2\over 3}\lambda^{3}
\eqno (3.6)
$$
$$
\zeta(0)_{\rm vector \; ghost \; modes}=-{13\over 90}
-2\lambda+{2\over 3}\lambda^{3}
\eqno (3.7)
$$
$$
\zeta(0)_{\rm decoupled \; ghost \; mode}={3\over 2} .
\eqno (3.8)
$$
By virtue of (3.1)-(3.8), the full $\zeta(0)$ value is
$$
\zeta_{\lambda}(0)
={89\over 90}- \lambda - 3\lambda^{2}
+{1\over 3} \lambda^{3} .
\eqno (3.9)
$$
Note that, since such a $\zeta(0)$ has a cubic dependence
on $\lambda$, a real value of $\lambda$ (compatible with
reality of the eigenvalues $E_{n}$ resulting from
self-adjointness) always exists such that our trace anomaly
(3.9) can agree with the values found in [4] in the case of
Luckock-Moss-Poletti boundary conditions:
$\zeta_{\rm LMP}(0)=-{758\over 45}$, or Barvinsky
boundary conditions: $\zeta_{\rm B}(0)=-{241\over 90}$.
More precisely, since a third-order algebraic equation with
real coefficients always admits at least one real solution,
we know a priori that the equations
$\zeta_{\lambda}(0)=-{758\over 45}$, or
$\zeta_{\lambda}(0)=-{241\over 90}$, are solved by at least
one real value of $\lambda$. In the former case, one finds that
only one real value of $\lambda$ exists, and it lies in the
open interval $]2,3[$. In the latter case there are three
real roots: $\lambda_{1}=-3\sqrt{3}+4$,
$\lambda_{2}=3\sqrt{3}+4$, $\lambda_{3}=1$.
The remarkable agreement when $\lambda=1$
with the trace anomaly resulting from the Barvinsky boundary
conditions deserves further thinking.

Other relevant values of
$\zeta_{\lambda}(0)$ are the ones corresponding to the
gauge invariance of the boundary conditions (2.6) or (2.7).
By virtue of (3.9) one finds
$$
\zeta_{-1}(0)= -{121\over 90}
\eqno (3.10)
$$
$$
\zeta_{-2}(0)= -{1051\over 90}.
\eqno (3.11)
$$
As far as we can see,
our trace anomaly (3.9)
and the ones obtained in [4] reflect three different sets of
boundary conditions for Euclidean quantum gravity, i.e.
(1.1)-(1.5), (1.10)-(1.14) and (2.1)-(2.3), (2.6)-(2.7). The
experts may also find it useful to remark that, for $\lambda=0$,
the result (3.1) yields $\zeta_{\rm TT}(0)={112\over 45}$, which
agrees with section 7.3 of [2], which used a less powerful
algorithm.
\vskip 1cm
\leftline {\bf 4. Concluding remarks}
\vskip 1cm
\noindent
Although local supersymmetry makes it necessary to study
local boundary conditions involving complementary projection
operators which act on fields or potentials [2,7,10], the resulting
boundary conditions for metric perturbations are not invariant
under the whole set of transformations (1.8). In Euclidean
quantum gravity one has thus a choice between local boundary
conditions whose restrictions on the perturbed 3-metric are
not gauge-invariant, or other
local or non-local boundary conditions which
restrict perturbations of the 3-geometry in a gauge-invariant way.
Our paper has studied mixed boundary conditions in
Euclidean quantum gravity which involve, for the
perturbed 3-metric, the most general form of Robin boundary
conditions. The results of our investigation are as follows.

First, the boundary conditions of section 2 correspond to the
quantum boundary-value problem which admits as a particular case
the boundary conditions on the linearized magnetic curvature
studied in section 7.3 of [2],
and such that the boundary conditions on the
perturbed 3-metric are always gauge-invariant.
Second, a detailed evaluation
of one-loop amplitudes is indeed possible, and the resulting
trace anomaly, given by (3.9), differs in general from the $\zeta(0)$
values obtained from other local or non-local boundary conditions
studied in the literature [4-7,10]. Third, when the dimensionless
parameter $\lambda=1$, or $-3\sqrt{3}+4$, or $3\sqrt{3}+4$,
the trace anomalies resulting from the
boundary conditions (1.1)-(1.5) and (2.1)-(2.3), (2.6)-(2.7)
turn out to agree. Hence there seems to be a non-trivial
link between trace anomalies, when at least the boundary
conditions on the perturbed 3-geometry are gauge-invariant.

The result (3.9) is a relevant example of how quantum
amplitudes may depend on the boundary conditions.
However, the geometric form of the corresponding asymptotic
expansion of the heat kernel remains unknown (cf [4]).
So far, only the local boundary conditions (1.10)-(1.14)
make it possible to obtain both geometric and analytic
formulae for the trace anomaly. It also
appears interesting to investigate the non-local nature of
the one-loop effective action, which is a linear combination
of $\zeta(0)$ and $\zeta'(0)$, in the case of local and
non-local boundary conditions in Euclidean quantum gravity.
This calculation appears possible for the first time, after
the thorough analysis of scalar fields appearing in [15,16].
Moreover, it appears necessary to understand whether any set
of local and gauge-invariant boundary conditions can be found
in Euclidean quantum gravity when a 4-dimensional background
geometry is analyzed. In the two-dimensional case, the author
of [17] has found that for $R^{2}+T^{2}$ gravity with an
independent spin-connection, local and gauge-invariant
boundary conditions do actually exist. A possible extension
of such results to 4-dimensional Riemannian geometries with
boundary would be of crucial importance to complete the
analysis of mixed boundary conditions in Euclidean quantum gravity.
\vskip 1cm
\leftline {\bf Acknowledgments}
\vskip 1cm
\noindent
We are much indebted to Igor Mishakov and Giuseppe Pollifrone
for scientific collaboration on Euclidean quantum gravity,
whilst Andrei Barvinsky motivated our recent interest in
non-local boundary conditions. Correspondence with Dmitri
Vassilevich has been very helpful to improve our manuscript.
Our paper was supported in part by the European Union under
the Human Capital and Mobility Programme. Moreover, the research
described in this publication was made possible in part by Grant
No MAE300 from the International Science Foundation
and from the Russian Government. The work
of A Kamenshchik was partially supported by the Russian Foundation
for Fundamental Researches through grant No 94-02-03850-a,
and by the Russian Research Project ``Cosmomicrophysics".
\vskip 1cm
\leftline {\bf Appendix A}
\vskip 1cm
\noindent
The boundary conditions (1.1)-(1.5) are a particular case of
a more general set of mixed boundary conditions for
Euclidean quantum gravity [5]. The most general form of
mixed boundary conditions which are completely invariant
under the transformations (1.8) on metric perturbations
is [4]
$$
\Bigr[h_{ij}\Bigr]_{\partial M}
=\Bigr[{\widehat h}_{ij}\Bigr]_{\partial M}=0
\eqno (A.1)
$$
$$
\Bigr[\Phi_{\nu}(h)\Bigr]_{\partial M}
=\Bigr[\Phi_{\nu}(\widehat h)\Bigr]_{\partial M}=0
\eqno (A.2)
$$
$$
\Bigr[\varphi_{\nu}\Bigr]_{\partial M}=0
\eqno (A.3)
$$
where $\Phi_{\nu}$ is an {\it arbitrary} gauge-averaging
functional, and the conditions (A.3) result from
(A.1)-(A.2) and ensure their validity. The boundary
conditions (A.1)-(A.3) generalize to pure gravity the
magnetic boundary conditions for Euclidean Maxwell theory
[14], i.e.
$$
\Bigr[A_{k}\Bigr]_{\partial M}=0
\eqno (A.4)
$$
$$
\Bigr[\Phi(A)\Bigr]_{\partial M}=0
\eqno (A.5)
$$
$$
\Bigr[\varphi \Bigr]_{\partial M}=0
\eqno (A.6)
$$
where $A_{\mu}$ is the potential, $\Phi$ is an arbitrary
gauge-averaging functional, and $\varphi$ is the ghost
0-form.

The boundary conditions (2.1)-(2.3) and (2.6)-(2.7) are
instead a generalization of the electric boundary
conditions for Euclidean Maxwell theory [14], i.e.
$$
\Bigr[A_{0}\Bigr]_{\partial M}=0
\eqno (A.7)
$$
$$
\left[{\partial A_{k}\over \partial \tau}\right]_{\partial M}=0
\eqno (A.8)
$$
$$
\left[{\partial \varphi \over \partial \tau}
\right]_{\partial M}=0 .
\eqno (A.9)
$$
Note that (A.6) ensures the invariance of (A.4)-(A.5) under
the gauge transformations
$$
{\widehat A}_{\mu} \equiv A_{\mu}+\partial_{\mu}\varphi
\eqno (A.10)
$$
whilst (A.9) ensures the invariance of (A.7)-(A.8) under
the action of (A.10).
\vskip 1cm
\leftline {\bf Appendix B}
\vskip 1cm
\noindent
Since the calculations leading to (3.1)-(3.8) are not
straightforward, we give a brief outline of the most
difficult, i.e. the contribution of scalar-type
perturbations to the trace anomaly. To avoid repeating
ourselves, we refer the reader to [6,13,14] for the meaning
of our notation and for the technique used in our investigation.

As the eigenvalues tend to $\infty$ and the order $n$ of basis
functions tends to $\infty$, the $n$-dependent coefficient in
the equation obeyed by the eigenvalues by virtue of boundary
conditions takes the form (cf [4])
$$
\rho_{\infty}(n)=12{n(n^{2}-1)\over (n^{2}-4)} .
\eqno (B.1)
$$
Thus, since the degeneracy of coupled scalar modes is
$n^{2}$ for all $n \geq 3$, one has to consider
${1\over 2}n^{2}\log(\rho_{\infty}(n))$, whose expansion
as $n \rightarrow \infty$ does not have terms proportional
to ${1\over n}$. Hence $I_{\rm pole}(\infty)$ vanishes. Moreover,
as the eigenvalues tend to $0$ whilst $n \rightarrow \infty$,
the term contributing to $I_{\rm pole}(0)$ in the
eigenvalue condition reduces to (cf [4])
$$
\rho_{0}(n)=\Gamma^{-4}(n) \Bigr(1-{1\over n}\Bigr)
{(n^{2}+1)\over n(n+1)(n+2)}
\biggr[1+2{(\lambda+1)\over n}f_{1}(n)
+(\lambda+1)^{2}f_{2}(n)\biggr]
\eqno (B.2)
$$
where
$$
f_{1}(n) \equiv {(n^{4}-2n^{2}-1)\over (n^{2}+1)(n^{2}-4)}
\eqno (B.3)
$$
$$
f_{2}(n) \equiv {(n^{2}-1)\over (n^{2}-4)}
{1\over (n^{2}+1)} .
\eqno (B.4)
$$
Thus, as $n \rightarrow \infty$, ${1\over 2}n^{2}\log
(\rho_{0}(n))$ has many contributions proportional to
${1\over n}$, so that $I_{\rm pole}(0)$ is found to be
$$
I_{\rm pole}(0)=-{119\over 180}+\lambda
+{1\over 3}(\lambda+1)^{3} .
\eqno (B.5)
$$
Last, $I_{\rm log}$ is obtained after eliminating fake roots
in the eigenvalue condition, and then using the uniform
asymptotic expansion of Bessel functions. This leads to
$$
I_{\rm log}=18-{1\over 60} .
\eqno (B.6)
$$
Since $\zeta(0)=I_{\rm log}+I_{\rm pole}(\infty)
-I_{\rm pole}(0)$, equations (B.5)-(B.6) lead to the
result (3.2).
\vskip 1cm
\leftline {\bf References}
\vskip 1cm
\item {[1]}
Hawking S W 1984 {\it Nucl. Phys.} B {\bf 239} 257
\item {[2]}
Esposito G 1994 {\it Quantum Gravity, Quantum Cosmology
and Lorentzian Geometries} (Lecture Notes in Physics
{\bf m12}) (Berlin: Springer)
\item {[3]}
Hawking S W and Gibbons G W 1993 {\it Euclidean Quantum
Gravity} (Singapore: World Scientific)
\item {[4]}
Esposito G, Kamenshchik A Yu, Mishakov I V and Pollifrone
G {\it One-Loop Amplitudes in Euclidean Quantum Gravity}
(DSF preprint 95/16)
\item {[5]}
Barvinsky A O 1987 {\it Phys. Lett.} {\bf 195B} 344
\item {[6]}
Esposito G, Kamenshchik A Yu, Mishakov I V and Pollifrone
G 1994 {\it Phys. Rev.} D {\bf 50} 6329
\item {[7]}
Luckock H C 1991 {\it J. Math. Phys.} {\bf 32} 1755
\item {[8]}
York J W 1986 {\it Found. Phys.} {\bf 16} 249
\item {[9]}
Hartle J B and Hawking S W 1983 {\it Phys. Rev.}
D {\bf 28} 2960
\item {[10]}
Moss I G and Poletti S 1990 {\it Nucl. Phys.} B {\bf 341} 155
\item {[11]}
Schleich K 1985 {\it Phys. Rev.} D {\bf 32} 1889
\item {[12]}
Hawking S W 1977 {\it Commun. Math. Phys.} {\bf 55} 133
\item {[13]}
Barvinsky A O, Kamenshchik A Yu and Karmazin I P 1992
{\it Ann. Phys.}, {\it NY} {\bf 219} 201
\item {[14]}
Esposito G, Kamenshchik A Yu, Mishakov I V and Pollifrone G
1994 {\it Class. Quantum Grav.} {\bf 11} 2939
\item {[15]}
Dowker J S and Apps J S 1995 {\it Class. Quantum Grav.}
{\bf 12} 1363
\item {[16]}
Bordag M, Geyer B, Kirsten K and Elizalde E {\it Zeta-Function
Determinant of the Laplace Operator on the D-Dimensional
Ball} (UB-ECM-PF preprint 95/10)
\item {[17]}
Vassilevich D V {\it On Gauge-Invariant Boundary Conditions
for 2D Gravity with Dynamical Torsion} (TUW preprint 95/6)

\bye